\documentclass[a4paper,fleqn,usenatbib]{mnras}
\usepackage{newtxtext,newtxmath}
\usepackage[T1]{fontenc}
\usepackage{ae,aecompl}
\usepackage{array}
\usepackage[T1]{fontenc}
\usepackage{ae,aecompl}

\usepackage{graphicx}	% Including figure files
\usepackage{amsmath}	% Advanced maths commands
\usepackage{amssymb}	% Extra maths symbols
\usepackage{graphicx}	% Including figure files
\usepackage{float}
\usepackage{soul}

\title[GK Car and GZ Nor: Two low-luminous, depleted \\ RV Tauri stars]{GK Car and GZ Nor: Two low-luminous, depleted \\ RV Tauri stars}

\author[I. Gezer et al.]
   {I.~Gezer$^{1}$, H.~Van Winckel$^{2}$, R.~Manick$^{2}$, D.~Kamath$^{3,4}$ \\
  $^{1}$Nicolaus Copernicus Astronomical Center, Rabia\'{n}ska 8, 87-100 Toru\'{n}, Poland\\
  $^{2}$Institute of Astronomy, KU Leuven, Celestijnenlaan, 200D 3001 Leuven, Belgium\\
  $^{3}$Department of Physics and Astronomy, Macquarie University, Sydney, NSW 2100, Australia\\
  $^{4}$Australian Astronomical Observatory, PO Box 915, North Ryde, NSW 1670, Australia}

\date{Accepted XXX. Received YYY; in original form ZZZ}

\pubyear{2019}

\begin{document}
\label{firstpage}
\pagerange{\pageref{firstpage}--\pageref{lastpage}}
\maketitle

\begin{abstract}
We performed a photometric and spectroscopic analysis of two RV Tauri stars GK Car and GZ Nor. Both objects are surrounded by hot circumstellar dust.
Their pulsation periods, derived from ASAS photometric time series, have been used to derive their luminosities and distances via the PLC relation. 
In addition, for both objects, GAIA distances are available. The Gaia distances and luminosities are consistent with the values obtained from the PLC relationship. GK Car is at distance of 4.5$\pm{1.3}$ kpc and has a luminosity of 1520$\pm{840}$ L$_{\odot}$, while GZ Nor is at distance of 8.4$\pm{2.3}$ kpc 
and has a luminosity of 1240 $\pm{690}$ L$_{\odot}$. Our abundance analysis reveals that both stars show depletion of refractory elements with [Fe/H]=$-$1.3 and
[Zn/Ti]=$+$1.2 for GK Car and [Fe/H]=$-$2.0 and [Zn/Ti]=$+$0.8 for GZ Nor. In the WISE colour-colour diagram, GK Car is located in the RV Tauri box
as originally defined by \cite{evans85} and updated by \cite{gezer15}, while GZ Nor is not. Despite this, we argue that both objects are surrounded by a gravitationally bound disc. As depletion is observed in binaries, we postulate that both stars are binaries as well. RV Tauri stars are generally acknowledged to be post$-$AGB stars. Recent studies show that they might be either indeed post$-$AGB or post$-$RGB objects depending on their luminosity. For both objects, the derived luminosities are relatively low for post-AGB objects, however, the uncertainties are quite large. We conclude that they could be either post-RGB or post-AGB objects.

\end{abstract}

\begin{keywords}
stars: abundances -- stars: AGB and post-AGB -- stars: evolution -- stars: individual: GK Car -- stars: individual: GZ Nor -- stars: individual: RV Tauri
\end{keywords}

%%%%%%%%%%%%%%%%%%%%%%%%%%%%%%%%%%%%%%%%%%%%%%%%%%

%%%%%%%%%%%%%%%%% BODY OF PAPER %%%%%%%%%%%%%%%%%%

\section{Introduction}

RV Tauri stars are population II Cepheid variables with spectral types typically between F and K. There are 126 RV Tauri stars known in the Galaxy (see GCVS \cite{samus09}) and many of them have been discovered in the Small Magellanic Cloud (SMC) and Large Magellanic Cloud (LMC) \citep{alcock98,soszynski08,buchler09,soszynski10}. They owe their name to the prototype RV Tau. A defining characteristic of RV Tauri stars is the presence of subsequent deep and shallow minima in their light curves \citep[e.g.][]{pollard97}. The period between two 
successive deep minima is called the \textquotedblleft formal\textquotedblright{} period and lies in the range 20-150 days. The period between a deep and shallow minimum is typically half of the
formal period and is called the \textquotedblleft fundamental\textquotedblright{} period \citep{soszynski08, soszynski10}. There are several hypotheses for their pulsation mechanism 
\citep{tuchman93}. Non-linear, non-adiabatic hydrodynamical RV Tauri models show that the alternating minima in the light curves might be explained with double mode pulsations between the
fundamental mode and the first overtone \citep{fokin94}. In addition, it can also be explained using chaotic systems \citep{buchlerandkovacs87}.

IRAS (Infrared Astronomical Satellite 1983) detected several RV Tauri stars which show large infrared excesses (IR) due to thermal emission from dust. On the basis of this finding, 
RV Tauri stars were classified as post-Asymptotic Giant Branch (post$-$AGB) objects by \cite{jura86}. As post-AGB stars, they are expected to display AGB nucleosynthesis products, which are mostly 
C and {$\it{s}$}$-$process elements. However, most RV Tauri stars that have been chemically studied to date show no traces of C and {$\it{s}$}$-$process enhancements, with the exceptions of HD 158616 \citep{winckel97}, MACHO 47.2496.8 \citep{reyniers07a} and SMC$-$T2CEP$-$018 (SMC J005107.19$-$734133.3) \citep{kamath14}. Instead, RV Tauri stars show often a chemical anomaly which is 
called \textquotedblleft depletion\textquotedblright{} \citep[e.g.][]{giridhar94,giridhar98,giridhar00,gonzalez97b,gonzalez97a, vanwinckel98, maas02,maas03,maas05,deroo05a, gezer15}. 
In a depleted photosphere, refractory elements, which have high dust condensation temperature, are underabundant, while volatiles, which has low dust condensation temperature, are more abundant 
\citep[e.g.][]{vanwinckel03}. It is assumed that the dust formation causes chemical fractionation in the circumstellar environment and the radiation pressure on dust grains separates the dust from the gas. 
The cleaned gas is re-accreted onto the stellar surface. Such a process seems to occur only in a stable circumstellar disc \citep{waters92}. The most characteristic chemical signatures of depleted photospheres 
are high [Zn/Fe], [Zn/Ti] and [S/Ti] ratios, which are used to identify a photosphere to be depleted or not \citep{gezer15}.

\begin{figure}
\begin{center}
\includegraphics[scale=0.23]{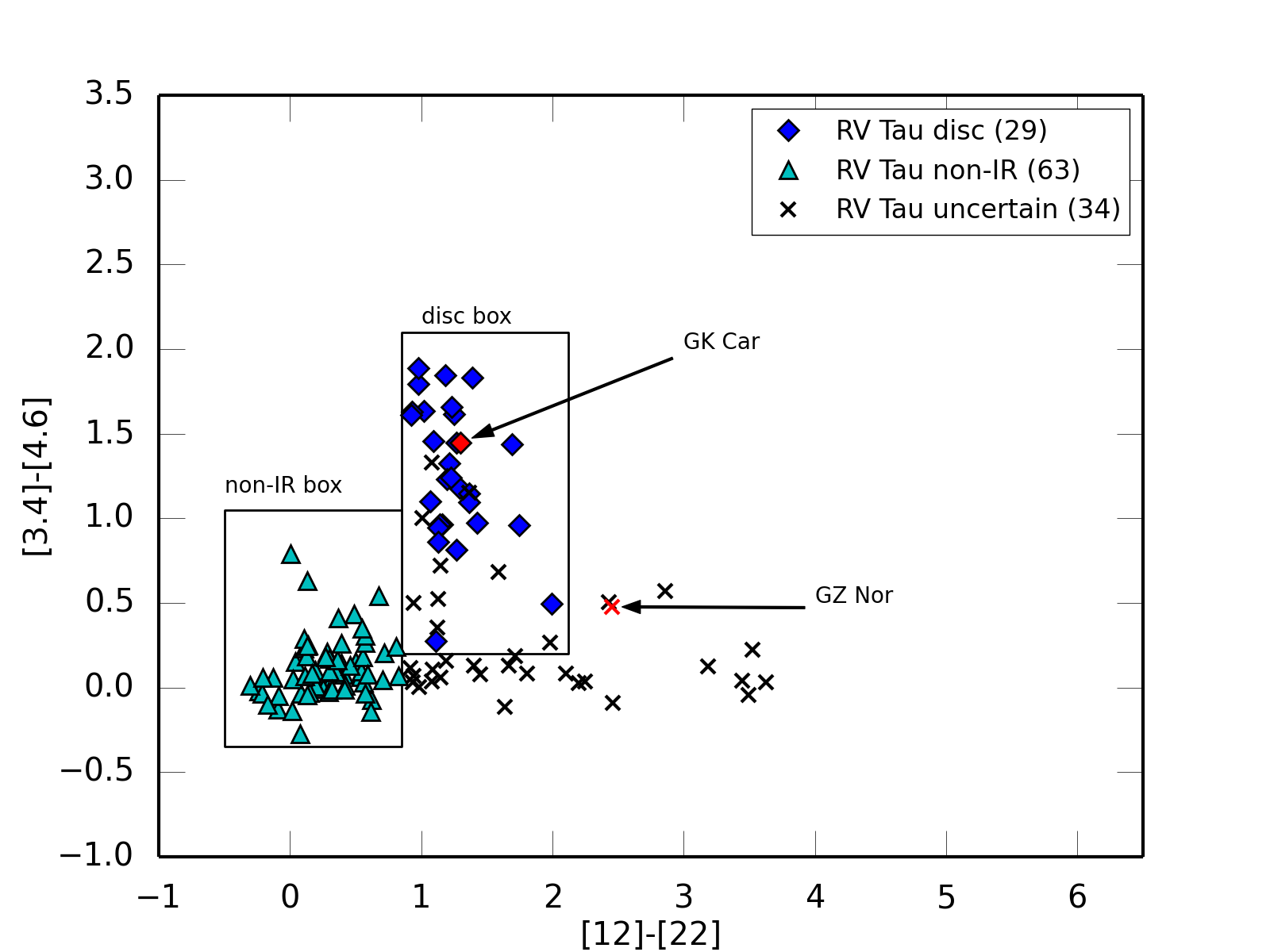}
\caption{WISE colour-colour diagram for the Galactic RV Tauri stars. Different type of SED characteristic occupy different region in the diagram.}\label{fig_1} 
\end{center}
\end{figure}

\cite{evans85} showed that IR colours of RV Tauri stars occupy a specific region in the IRAS [12]-[25], [25]-[60] diagram and he defined this region as the \textit{RV Tauri box}. 
In our first paper (see \cite{gezer15}, hereafter Paper I), we have expanded his study using WISE (Wide-Field Infrared Survey) data, which is similar to IRAS but deep enough to detect all Galactic 
RV Tauri stars. We defined a new WISE colour combination as [3.4]-[4.6] versus [12]-[22], which is a good WISE alternative of IRAS colour-colour diagram. We also performed a detailed spectral energy 
distribution (SED) analysis for all Galactic RV Tauri stars and showed that they display three different types of SED characteristics; \textit{disc}, \textit{non-IR} and \textit{uncertain} types. This classification 
is shown in  Fig. \ref{fig_1}. A disc source displays a very distinctive SED with the IR-excess starting in the near-IR region and peaking around 10 $\mu$m. This type of SED is a clear indication of the 
presence of a stable compact dusty disc \citep{deruyter06, deroo06, deroo07, gielen11, hillen15}. Disc-type SEDs occupy a specific region in the WISE colour-colour diagram and this region is equivalent 
to the \textit{RV Tauri box} defined by \cite{evans85} in the IRAS colour-colour diagram. Thus, we slightly redefined the \textit{RV Tauri box} as a \textit{disc box} in the WISE colour-colour diagram. 
RV Tauri stars which do not show any infrared excess, cluster around the zero-point of the diagram, and we define this region as the \textit{non-IR box}. For the remaining RV Tauri stars, the full SED is not 
clear, which is likely because of the large amplitude pulsations or/and a poor photometric sampling. Besides, these sources are mostly located outside of the \textit{disc} and \textit{non-IR} boxes in the 
WISE colour-colour diagram. Therefore, we classify these sources as \textit{uncertain}.

Recently, \cite{manick18} also showed that the SMC and LMC RV Tauri stars display similar IR characteristics as their Galactic counterparts. The authors performed a systematic study of RV Tauri stars 
in the SMC/LMC and they yield their luminosities using the known distance to the Magellanic Clouds. They also interpret the evolutionary nature of RV Tauri stars using
their luminosities. Objects which display disc-type SEDs are very likely binaries and they have a wide range of luminosities. Luminous binaries are likely post$-$AGB objects while the 
low luminous ones are probably post-red giant branch (post-RGB) objects. Post-RGB objects are a new class of dusty objects discovered in SMC/LMC surveys \citep{kamath14, kamath15}. These objects 
are similar to post$-$AGB objects in terms of their stellar parameters and infrared excess, except that they display lower luminosities (100-2500 L$_{\odot}$) than post$-$AGB stars. These stars very likely
evolve off the RGB because of a strong binary interaction process already occurring on the RGB \citep{kamath16}.

In this paper, we selected two as yet poorly studied RV Tauri stars which have different SED characteristics and locations in the WISE colour-colour diagram. The one is a disc source GK Car 
which is located in the \textit{disc box} and another one is GZ Nor which is outside of the \textit{disc box} and defined as \textit{uncertain}. Their positions in the WISE colour-colour diagram are shown in Fig. \ref{fig_1}. The outline of the paper is as follows: Photometric and spectroscopic data and SEDs are presented in Section 2. Section 3 describes the photometric analysis including the pulsations and determination
of luminosity and distance based on two methods; the Period-Luminosity-Distance (PLC) relationship and GAIA. Determination of atmospheric parameters and abundance analysis of the stars 
are presented in Section 4. In Section 5 conclusions are summarized.

\section{Data}

Basic parameters, WISE colours and SED types of the stars are given in Table \ref{table_1}.

\begin{table*}
 \centering
  \caption{Basic parameters of GK Car and GZ Nor are from SIMBAD. WISE colours and SED types are from our previous study \citep{gezer15}.}
  \begin{tabular}{lccccccl}
  \hline
   Name & \multicolumn{1}{p{2cm}}{\centering R.A. \\ (J2000)} & \multicolumn{1}{p{2cm}}{\centering Decl. \\ (J2000)}& l& b& [12]$-$[22]&[3.4]$-$[4.6]&SED  \\
  \hline

  GK Car & 11 14 01.61& $-$57 43 15 &290.1955 &$+$02.7218 & 1.301 & 1.448 & Disc\\
  
  GZ Nor & 16 31 54.15& $-$55 33 07& 330.6957& $-$05.0546 &  2.453 & 0.477 & Uncertain\\ 
\hline					        						   	  	  
\end{tabular}
\label{table_1}
\end{table*}

\subsection{Photometric data}
\subsubsection{ASAS photometry}

We used data from the All Sky Automated Survey (ASAS, \cite{pojmanski02}) for the photometric analysis of GK Car and GZ Nor. ASAS itself is an ongoing project to detect any kind 
of photometric variability and produce extensive catalogues of variable stars (ACVS) by monitoring the large area of the sky with fully automated instruments. The prototype of the project 
was first operated in 1996 at the Warsaw University Astronomical Observatory. Now, it carries on with three fully automatic instruments having $V$ and $I$ filters attached to 2K and 4K CCD 
cameras at Las Campanas Observatory in Chile and at Mt. Haleakala Observatory in Maui, Hawaii. Through the project, nearly 20 million stars which are optically brighter than 14 magnitudes 
have been observed so far. The photometric accuracy is about 0.05 mag. 

\begin{figure*}
\begin{center}
\begin{tabular}{c}
\includegraphics[width=15cm]{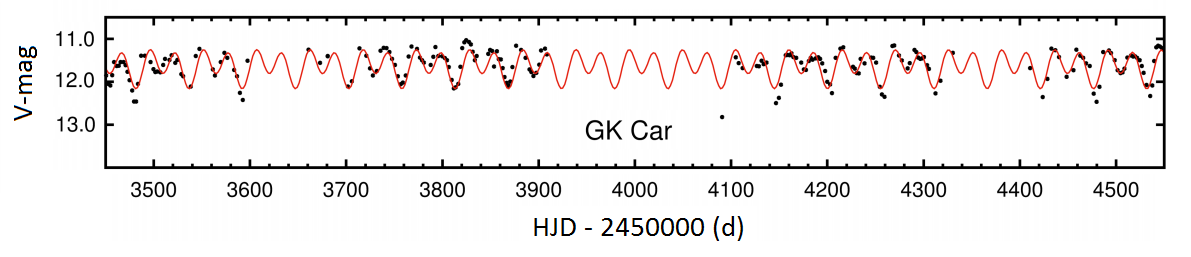}\\
\includegraphics[width=15cm]{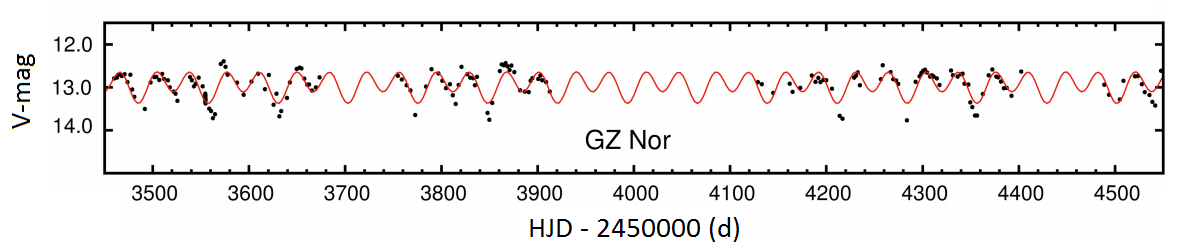}
\end{tabular}
\caption{Time-series of the pulsations obtained using ASAS and AAVSO photometric data of GK Car and GZ Nor are shown here.
Red line represents the calculated light curve and the black dots are observation points.}\label{fig_2}. 
\end{center}
\end{figure*}

The public domain data were taken from the AASC\footnote{http://www.astrouw.edu.pl/asas/?page=acvs} and ACVS\footnote{http://www.astrouw.edu.pl/asas/?page=aasc} websites. For both 
stars, we use the ASAS-3 configuration with a wide-field 2Kx2K CCD camera. ASAS is currently using 5 different sizes for the aperture photometry, which ranges from 2 pixels to 6 pixels and 
recommends using an aperture size according to the V-magnitude of the star. We used the smallest, 2 pixels wide aperture size, which is recommended by the ASAS guidelines 
for stars fainter than 12 magnitudes. We also limited our photometric analysis with the A-grade data quality. The time-series of the photometric pulsations of GK Car and GZ Nor are shown in 
Fig.\ref{fig_2}. Photometric information of the stars taken from ASAS is given Table \ref{table_2}.

\begin{table}
%  \setlength{\tabcolsep}{1pt}
%    \scriptsize
    \caption{Photometry info. No. refers to number of observational points.}
      \begin{tabular}{lcccc}
  \hline
   Name  & m$_{v}$ & m$_{I}$ &  No. & $\Delta{T}$ (day)  \\
  \hline

  GK Car  & 11$^m$.34 & 10$^m$.15 & 530 &  3283 \\
 
  GZ Nor  & 13$^m$.13 & 11$^m$.64 & 417 &  3166 \\
\hline                                                                  
\end{tabular}
\label{table_2}
\end{table}

\subsection{Spectroscopic data}
\subsubsection{UVES spectra}

\begin{figure}
\begin{center}
%\begin{tabular}{c}
\includegraphics[scale=0.23]{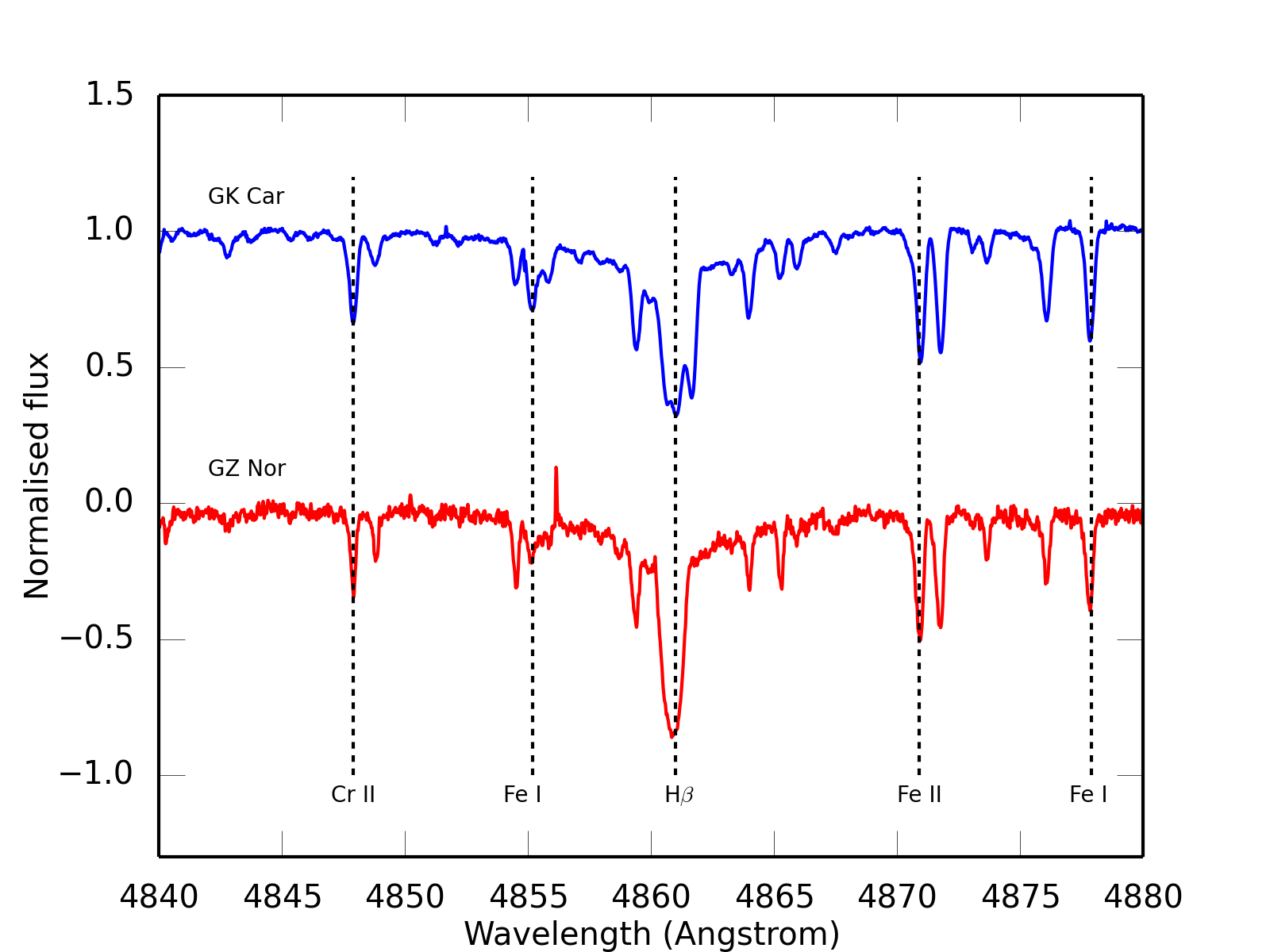}
%\end{tabular}{c}
\caption{The normalised UVES spectra of GK Car and GZ Nor around the H$_{\beta}$ region.}\label{fig_3}
\end{center} 
\end{figure}

High-resolution, high signal-to-noise spectra for GK Car and GZ Nor were obtained with the Ultraviolet and Visual Echelle Spectrograph (UVES; \cite{dekker00}), mounted on the 8m UT2 Kueyen
Telescope of the VLT array at the Paranal Observatory of ESO in Chili. The dichroic beam splitter provides a wavelength coverage from approximately 3760 to 4985 \AA{} for the blue arm, 
6705 to 8513 \AA{} and 8663 to 10420 \AA{} for the lower and upper parts of the mosaic CCD chip in the red arm, respectively. Observations were made in 2005 within the scope of 074.D-0619 
ESO program. The log of the observations is given in Table \ref{table_3}. The data were reduced using the pipeline reduction in the MIDAS environment. The reduction includes extraction of 
spectral orders, wavelength calibration, applying this wavelength scale to flat-field divided data and cosmic-clipping. We normalised the spectra using fifth order polynomials 
on continuum points indicated interactively. A small part of the normalised spectra of GK Car and GZ Nor are shown in Fig. \ref{fig_3}.

\begin{table}
  \scriptsize
    \caption{Observational logs of the stars.}
     \begin{tabular}{lcccc}

\hline
   Name & Date &  UT  &  Exp.time &  r$_{v}$ \\
 & & (start) & (s) & (km/s) \\
\hline
  GK Car &2005/02/09&  08:03  & \multicolumn{1}{p{2cm}}{\centering
Blue: 1x1800 \\ Red: 1x1800} & $-$22.3 $\pm$0.2  \\
  GZ Nor &2005/02/09& 08:38  & \multicolumn{1}{p{2cm}}{\centering
Blue: 1x1800 \\ Red: 1x1800}& $-$123.8 $\pm$0.1 \\
\hline
\end{tabular}
\label{table_3}
\end{table}

\subsection{Spectral energy distribution (SED)}\label{SED_section}

For both GK Car and GZ Nor, full SEDs are constructed from the available photometric data in the Vizier database \citep{vizier00}. Scaled Kurucz models \citep{kurucz03} 
were used with stellar parameters T$_{\rm eff}$, log$g$ and [Fe/H] obtained during our atmospheric parameters determination (see section \ref{atmospar}). In order to 
reach the optimised Kurucz model, the SED fitting is done by using a grid-method explained in \cite{degroote11}. 

\begin{figure}
\begin{center}
%\begin{tabular}{c}
\includegraphics[scale=0.58]{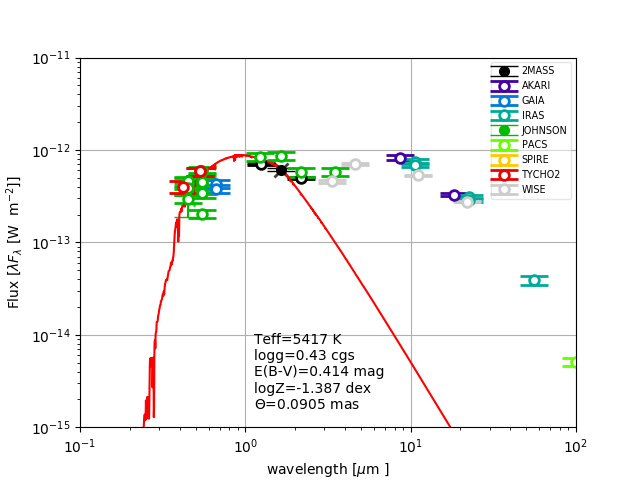}\\
\includegraphics[scale=0.58]{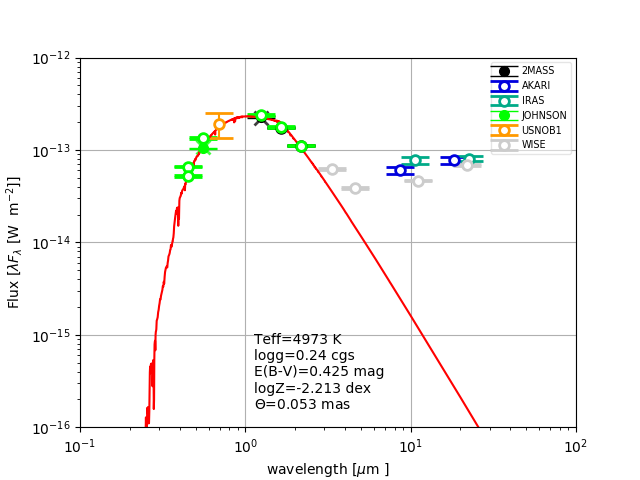}
%\end{tabular}
\caption{The SED of GK Car is shown in the top panel. The bottom panel shows the SED of GZ Nor.}\label{fig_4}
\end{center}
\end{figure}

The total reddening of E(B$-$V) is determined by minimising the difference between the dereddened photometric data and the appropriate Kurucz model atmosphere. We assume that the total 
reddening includes both the interstellar and circumstellar reddening. We also assume that the circumstellar component follows the interstellar extinction law. The SEDs are displayed in Figure 
\ref{fig_4}. For GK Car, we obtained a total reddening of E(B$-$V)=0.4$\pm$0.1, while for GZ Nor a total reddening of E(B$-$V)=0.4$\pm$0.1 was found. 

\section{Photometric Analysis and the PLC relation}
\subsection{Pulsation analysis}

The raw photometric data were first checked for outliers beyond 4 $\sigma$ from the mean. We then used this cleaned photometric data for our pulsation analysis. The period analysis has been
performed using Period04 \citep{lenz05}, which uses the Discrete Fourier Transform to find the most dominant frequency of a light curve. After the most dominant frequency is obtained, it is 
prewhitened and the second most dominant frequency in the remaining data is searched for. This process is repeated until the signal-to-noise (S/N) ratio falls below 4. This critical S/N value $\thicksim$ 
4 is used, because frequencies with a lower S/N cannot be distinguished from noise \citep{breger93, kuschnig97}.

\begin{figure}
\begin{center}
%\begin{tabular}{c}
\includegraphics[scale=0.23]{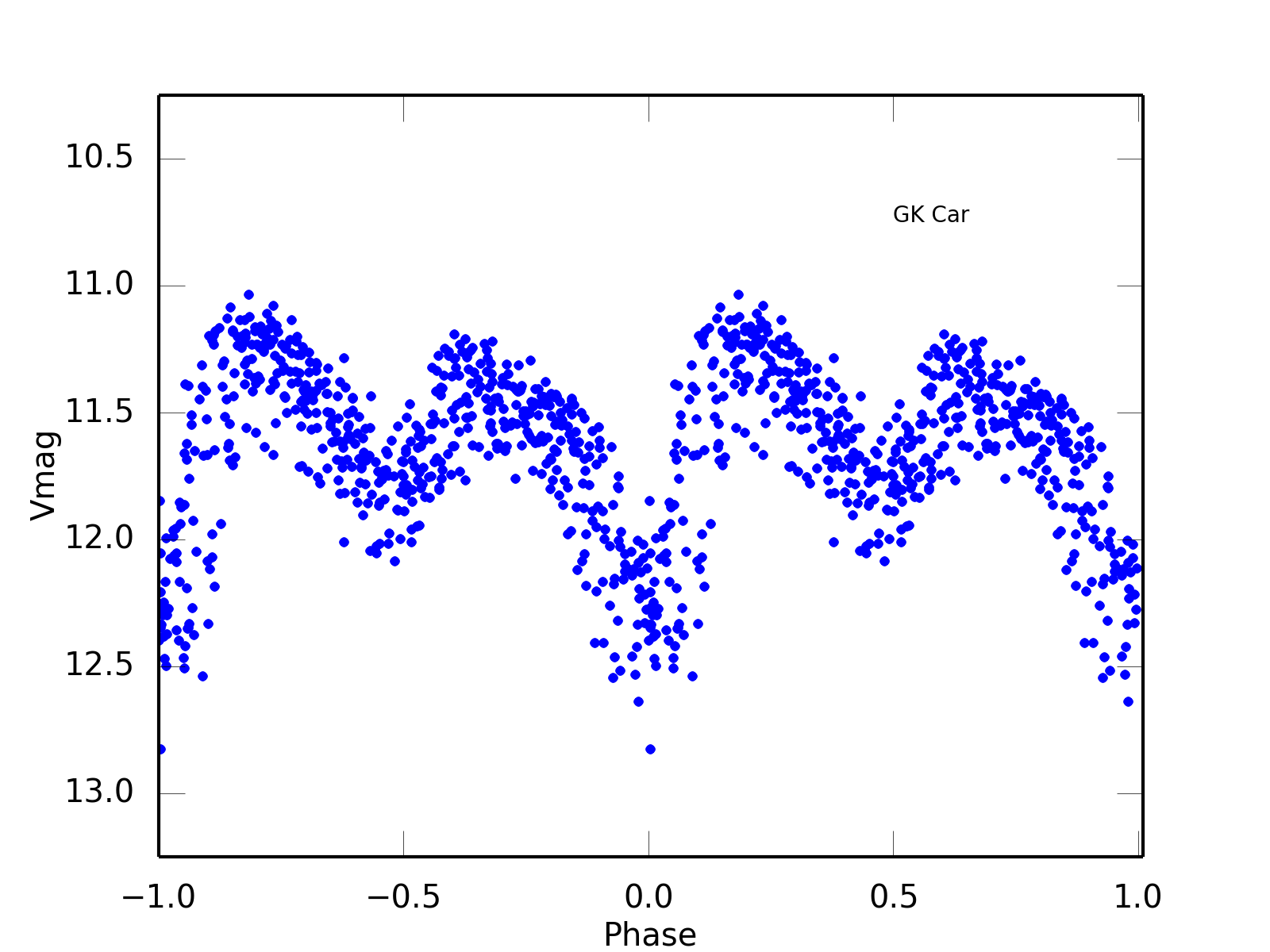}\\
\includegraphics[scale=0.23]{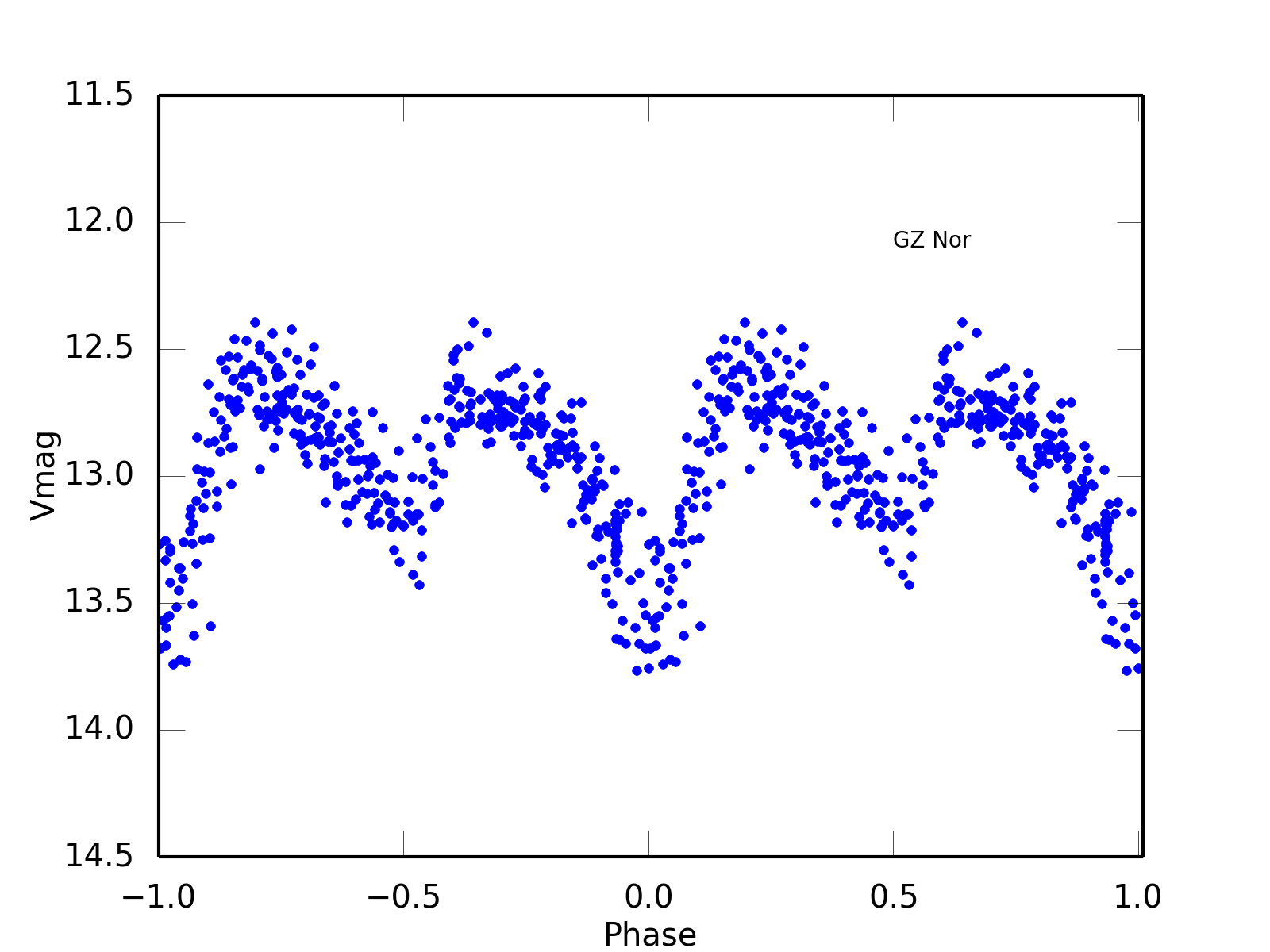}
%\end{tabular}
\caption{The phase plot of GK Car folded on a period of 55.2 days. GZ Nor phase plot folded on a period of 72.4 days.}\label{fig_5}. 
\end{center}
\end{figure}

For GK Car, we obtained two periods, a fundamental period of 27.6 days together with a 55.2 days formal period, both are consistent with the literature. GK Car was first identified 
in \cite{rosino51} as an RV Tau variable with a 55.6 days pulsation period. It is also given in the GCVS \citep{samus09} with a 55.3 days period and a RVa photometric class. The phased 
photometric data for GK Car is shown in the top panel of Fig. \ref{fig_5}, folded on the formal period. 

GZ Nor is given as a long period variable star in GCVS with a 35.9 days period based on \cite{kruytbosch32}. Using the more recent ASAS photometry we detected two periods above S/N 4, a 
fundamental period of 36.2 days together with a 72.4 days formal period. A phase plot of photomotric data of GZ Nor folded on the formal period is shown in Fig. \ref{fig_5}. 
It is clearly seen that it shows subsequent deep and shallow minima, which is a characteristic of RV Tau variables. We couldn't detect any variability in the mean magnitude. Thus, we classified GZ Nor 
as an RVa photometric type object \citep{kukarkin58}.

\subsection{Luminosity and Distance Estimates using the PLC relation}\label{luminosity-distance}

Luminosity is an important parameter and often difficult to obtain. Deriving the luminosities of type II Cepheids is especially important in the context of the direct relationship 
between the period of pulsation and the luminosity. Some period-luminosity (P-L) relations for type II Cepheids in the galactic globular clusters, SMC and LMC are given by 
\cite{nemec94}, \cite{alcock98} and \cite{ripepi15}. More recently, \cite{manick17} gave the period-luminosity-colour (PLC) relation based on 187 LMC type II Cepheids. Here 
we use the following,

\begin{equation}\label{eq1}
% \begin{split}
M_{\rm v} = -2.53(\pm{0.03})\log{P}-1.17(\pm{0.03})+2.55\langle(V-I)_{\rm 0}\rangle \\ 
% \end{split}
\end{equation}\\

Although ASAS provides $V$ and $I$ band photometry for many objects, unfortunately only $V$ band photometry was available for our stars in the ASAS database.
We used the mean observed ASAS $V$ magnitudes and synthetic $I$ magnitudes which are calculated from the stellar models. We defined a zero point for this synthetic system using
the flux around the $I$ band and we calculated $I$ magnitudes for each star. For GK Car the obtained I magnitude is 10.15 mag and for GZ Nor it is 11.64 mag. In order to obtain intrinsic
(V-I)$_{0}$ from the observed (V-I), we used the relation E(V-I)=1.38$\times$E(B$-$V) from \cite{tammann03}. The E(B$-$V) values are obtained from the SED fits. 

\begin{table}
\scriptsize
 \caption{The fundamental pulsation period (P$_{0}$) is given in
Col.2. Col. 3 is the intrinsic (V$-$I)$_{0}$  values. Col. 4 is the derived E(B-V) values towards each object
obtained from the reddened SED model. Calculated distances using PLC relation are shown in Cols. 5 and luminosities which are
calculated using the distances in Col. 5 are given in Col.6.}
 \begin{tabular}{lccccc}
\hline
Star  & P$_{puls}$ & (V-I)$_{0}$ &  E(B-V) & Distance$_{PLC}$ & L$_{SED}$   \\
        & (days) & &  & (kpc)&  (L$_{\odot}$) \\
\hline
GK Car      &    27.6, 55.2 & 0.624  & 0.41$\pm{0.1}$ &4.54$\pm${1.26} &1520$\pm{840}$  \\
GZ Nor      &    36.2, 72.4 & 0.910  & 0.42$\pm{0.1}$ &8.38$\pm${2.32} &1240$\pm{690}$ \\
\hline
\end{tabular}
\label{table_4}
\end{table}

Once (V-I)$_{0}$ values are obtained for both stars, it is easy to derive luminosities using the absolute magnitudes from the PLC relationship. The relation \ref{eq1} turns 
M$_{V}$ = -3.22 mag and M$_{V}$ = -2.79 mag for GK Car and GZ Nor, respectively. We derived the reddening-corrected distances for our both stars. The amount of extinction, A$_{v}$, 
towards the star is E(B$-$V)$\times$R$_{v}$. We used R$_{v}$ = 3.1 for the total to selective Galactic extinction \citep{weingartner01} and used the E(B$-$V) values for each star from 
the SED fitting. For GK Car the total extinction A$_{v}$ is 1.27 and for GZ Nor it is 1.30. The obtained distances are given in column 5 of Table \ref{table_4}. We then computed the bolometric 
luminosities, L$_{SED}$, for each star using the integrated flux below the SED photospheric model scaled to the dereddened photometry using the obtained distances. The obtained SED luminosities 
are given in column 6 of Table \ref{table_4}.

\begin{table}
\scriptsize
 \caption{The GAIA distance is given in Col.2, lower and upper limits in the distances are given in Col.3 and 4, respectively \citep{bailer_jones18}. Col.5, 6 and 7 are luminosities, which are 
calculated using the distances in Col.2, 3 and 4, respectively.}
 \begin{tabular}{lcccccc}
\hline
Star  & Distance & Distance &  Distance & L$_{SED}$&  L$_{SED}$ & L$_{SED}$ \\
        &            &(lower)  & (upper)  &     & (lower)& (upper) \\
        &   (kpc)  &(kpc)  & (kpc)  & (L$_{\odot}$) & (L$_{\odot}$)& (L$_{\odot}$) \\
\hline
GK Car      & 3.782 &  3.376 & 4.293 &  1050 &837& 1352  \\
GZ Nor      & 8.800 &  6.630 & 12.062&1367  &  776 &2568\\
\hline
\end{tabular}
\label{table_5}
\end{table}

Recently, Gaia Data Release 2 (Gaia DR2) has become available through the Gaia Archive\footnote{https://www.cosmos.esa.int/gaia}, \citep{gaia18,luri18}. New Gaia DR2 
provides parallax for more than 1.3 billion objects. For many of the stars in GAIA DR2 data release, a reliable distance cannot be obtained by simply inverting the parallax. Therefore, for 1.3 
billion stars with parallaxes in the GAIA DR2 release, distances and their uncertainties are estimated from the parallax using a Galactic model as prior that varies as a function of Galactic longitude 
and latitude according to a Galaxy model \cite{bailer_jones15}. For our stars, the estimated GAIA distances with lower and upper limits are given in Table \ref{table_5}. We also calculated 
luminosities using the GAIA distances by integrating the SED model scaled to the dereddened photometry. The calculated luminosities with the related distances are given in column 5, 6 and 7 of 
Table \ref{table_5}, respectively.

\section{Chemical analysis}
\subsection{Atmospheric parameters}\label{atmospar}

To derive the photospheric parameters and elemental abundances on the basis of our high-resolution spectra, we used a python wrapper (PyMOOG) \citep{desmedt15} 
of the MOOG abundance code (version July 2009) \citep{sneden73} combined with  Kurucz-Castelli \citep{castelli04} and MARCS (Model Atmosphere in Radiative and Convective Scheme) 
\citep{gustafsson08} atmosphere models. The routine assumes the local thermal equilibrium (LTE) conditions. For spectral line identification, we used the Vienna atomic line database (VALD) 
\citep{kupka99} in combination with a line list compiled for the chemical analysis of A, F, and G stars at the Institute of Astronomy (KU Leuven) \citep{winckel00}.

The radial velocities are estimated for both stars using IRAF (Data Reduction and Analysis System) RV package by fitting a Gaussian curve to an identified atomic line. 
The radial velocity is found to be -22.3 $\pm{0.2}$ km/s for GK Car and -123.8 $\pm{0.1}$ km/s for GZ Nor, respectively. These radial velocities are accurate enough for line identification purposes.

Once lines are identified, the equivalent width (EW) are measured interactively by taking a direct integration in PyMOOG. Abundances are computed in an iterative process in which the theoretical EWs 
of single lines are computed for a given abundance and matched to the observed EWs. Line selection is quite important to determine accurate abundances. In order to limit ourselves to the linear part 
of the curve of growth, we select lines with EWs smaller than 150 m\AA{}. We also do not use lines with EWs smaller than 5 m\AA{} because they might be confused with noise in the spectra. Blended 
lines are also avoided. Synthetic spectra, which are modelled with MOOG using the VALD atomic data are used to check whether lines are blended with other identified lines.

It is crucial to define an accurate atmospheric model, which closely resembles the stellar effective temperature, surface gravity, metallicity and microturbulent velocity at the time of the spectral observation. 
These are derived using Fe I and Fe II lines. The effective temperature is obtained on the basis of the requirement that the abundances obtained from the Fe I lines are independent of the excitation potentials 
of the lines. The surface gravity is calculated based on the ionization balance between the Fe I and Fe II lines. The microturbulent velocity is calculated based on the assumption that the abundances obtained from 
the Fe I lines are independent of the reduced equivalent widths of the lines. 

For GK Car, using an (LTE) Kurucz-Castelli \citep{castelli04} model, we obtained T$_{\rm eff}$ = 5500 $\pm{125}$ K, log$g$ = 1.0 $\pm{0.25}$ dex, $\xi_{t}$ = 5.5 $\pm{0.5}$ km/s, and 
[FeI/H] = -1.32 $\pm{0.1}$. For GZ Nor, atmospheric parameters are calculated using both Kurucz-Castelli \citep{castelli04} and Marcs \citep{gustafsson08} models. GZ Nor is a cooler
star and we prefer to use the Marcs model in this case \citep{plez03}. For GZ Nor we obtained the following atmospheric parameters : T$_{\rm eff}$ = 4875 $\pm{125}$ K, log$g$ = 0.50 $\pm{0.25}$ dex, 
$\xi_{t}$ = 4.0 $\pm{0.5}$ km/s, [FeI/H] = -2.05 $\pm{0.098}$. 

\subsection{Abundances of GK Car}

\begin{figure}
\begin{center}
\includegraphics[scale=0.23]{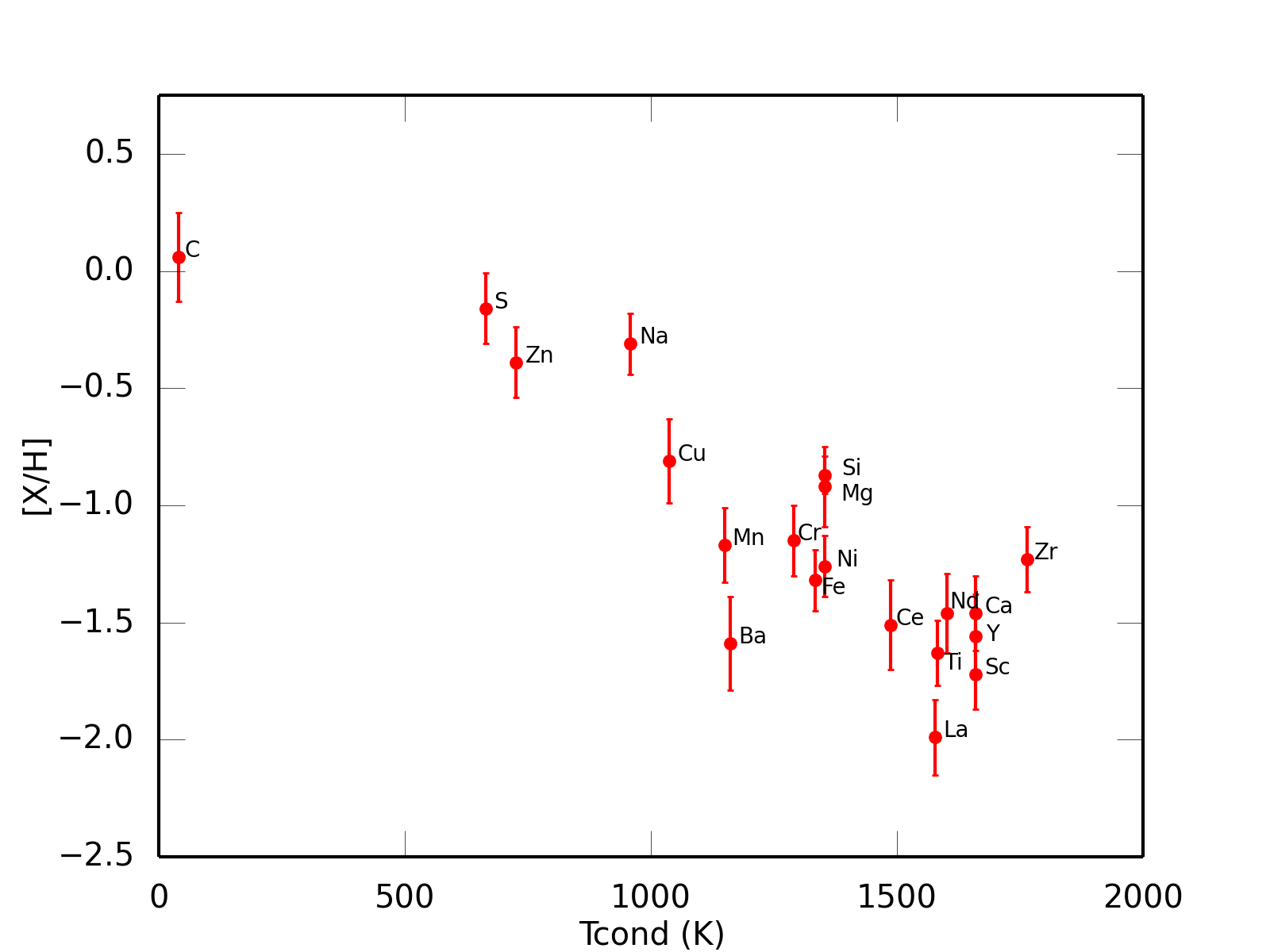}
\caption{The elements abundances in the atmosphere of GK Car versus the dust condensation temperature of elements.}\label{fig_6}  
\end{center}
\end{figure}

Owing to GK Car's high-resolution and high signal-to-noise spectra, a sufficient number (182 in total) of lines could be measured for our abundance analysis. 
The abundances are listed in Table \ref {table_6} and plotted versus their dust condensation temperatures in Fig. \ref{fig_6}. The small amount of line-to-line scatter, which is given in the $\sigma$ 
column in Table \ref{table_6} indicates the resulting abundances derived from the different lines of each element are consistent.

S and Zn abundances are good indicators of whether the stellar atmosphere is affected by the depletion process or not. Zn has a similar nucleosynthetic origin to Fe but it has a much lower condensation 
temperature. Similarly, S is a $\alpha$ element which has a lower condensation temperature than other $\alpha$ elements such as Mg, Ca and Ti. Hence, the low Fe ([Fe/H]=-1.3) abundance combined 
with the high Zn abundance ([Zn/Fe] = +0.9) and also the high S abundance ([S/Fe] = +1.2) versus the average abundance of the other $\alpha$ elements ([$\alpha$/Fe] = 0) shows that the photospheric 
abundances of GK Car are altered by the depletion process. In figure \ref{fig_6}, we display a clear correlation between the photospheric abundances and the condensation temperature.

The abundance ratios such as [Zn/Ti] and [S/Ti] are less dependent on the model atmospheric parameters than on the absolute abundance values. When the model temperature is 
increased by 125 K, Ti abundance will change about 0.14 dex but Zn abundance will also change about 0.10. In this case, the [Zn/Ti] ratio obtained from the hotter model is about the 
same ratio obtained from the model that we use. Relative abundance ratios are more reliable and less dependent on model parameters errors.

For several \textit {s}$-$process elements such as Y, Zr, Ba, La, Ce and Nd we could determine abundances. \textit {s}$-$process elements generally have high condensation temperatures and we found 
that abundances of \textit {s}$-$process elements in GK Cars photosphere follow indeed the depletion of Fe (see Table \ref{table_6}).

\begin{table}
 \centering
  \caption{Abundance analysis of GK Car. The solar abundances to compute the [X/H] ratios are taken from \protect\cite{asplund09}.
  The dust condensation temperatures, which is the temperature at which half of the particles of the element are condensed
  onto dust grains, under the assumption of chemical equilibrium, at constant pressure of 10$^{-4}$ bar in a solar mixture, 
  are from \protect\cite{lodders09}.}
  \begin{tabular}{lcccc}
  \hline
  & & \multicolumn{1}{c}{GK Car}& & \\
  & & \multicolumn{1}{c}{T$_{eff}$ $=$ 5500 K}& & \\
  & & \multicolumn{1}{c}{log$g$ $=$ 1.0} & & \\
  & & \multicolumn{1}{c}{$\xi_{t}$ $=$ 5.5 km/s}& & \\
  & & \multicolumn{1}{c}{[Fe/H] $=$ $-$1.32}& & \\
  \hline
   ion     &   N & [X/H]    & $\sigma$ & T$_{cond}$    \\
  \hline

CI   &5  &   0.06    & 0.16& 40   \\
NaI  &4  &  $-$0.31    & 0.15& 958  \\
MgI  &2  &  $-$0.92    & 0.18& 1354 \\
SiI  &3  &  $-$0.87    & 0.03& 1354 \\
SI   &3  &  $-$0.16    & 0.1& 664  \\
CaI  &7  &  $-$1.46    & 0.20& 1659 \\
ScII &4  &  $-$1.72    & 0.07& 1659 \\
TiII &15 &  $-$1.63    & 0.14& 1582 \\
CrII &8  &  $-$1.15    & 0.17& 1291 \\
MnI  &6  &  $-$1.17    & 0.09& 1150 \\
FeI  &77 &  $-$1.32    & 0.11& 1334 \\
FeII &8  &  $-$1.3     & 0.09&     \\
NiI  &15 &  $-$1.26    & 0.08& 1353 \\
CuI  &3  &  $-$0.81    & 0.13& 1037\\
ZnI  &3  &  $-$0.39    & 0.14& 726 \\
YII  &3  &  $-$1.56    & 0.16& 1659  \\
ZrII &3  &  $-$1.23    & 0.05& 1764 \\
BaII &2  &  $-$1.59    & 0.17& 1162 \\
LaII &2  &  $-$1.99    & 0.05& 1578 \\
CeII &5  &  $-$1.51    & 0.18& 1487 \\
NdII &4  &  $-$1.46    & 0.1& 1602 \\
\hline					        						   	  	  
\end{tabular}
\label{table_6}
\end{table}

\subsection{Abundances of GZ Nor}

\begin{figure}
\begin{center}
\includegraphics[scale=0.23]{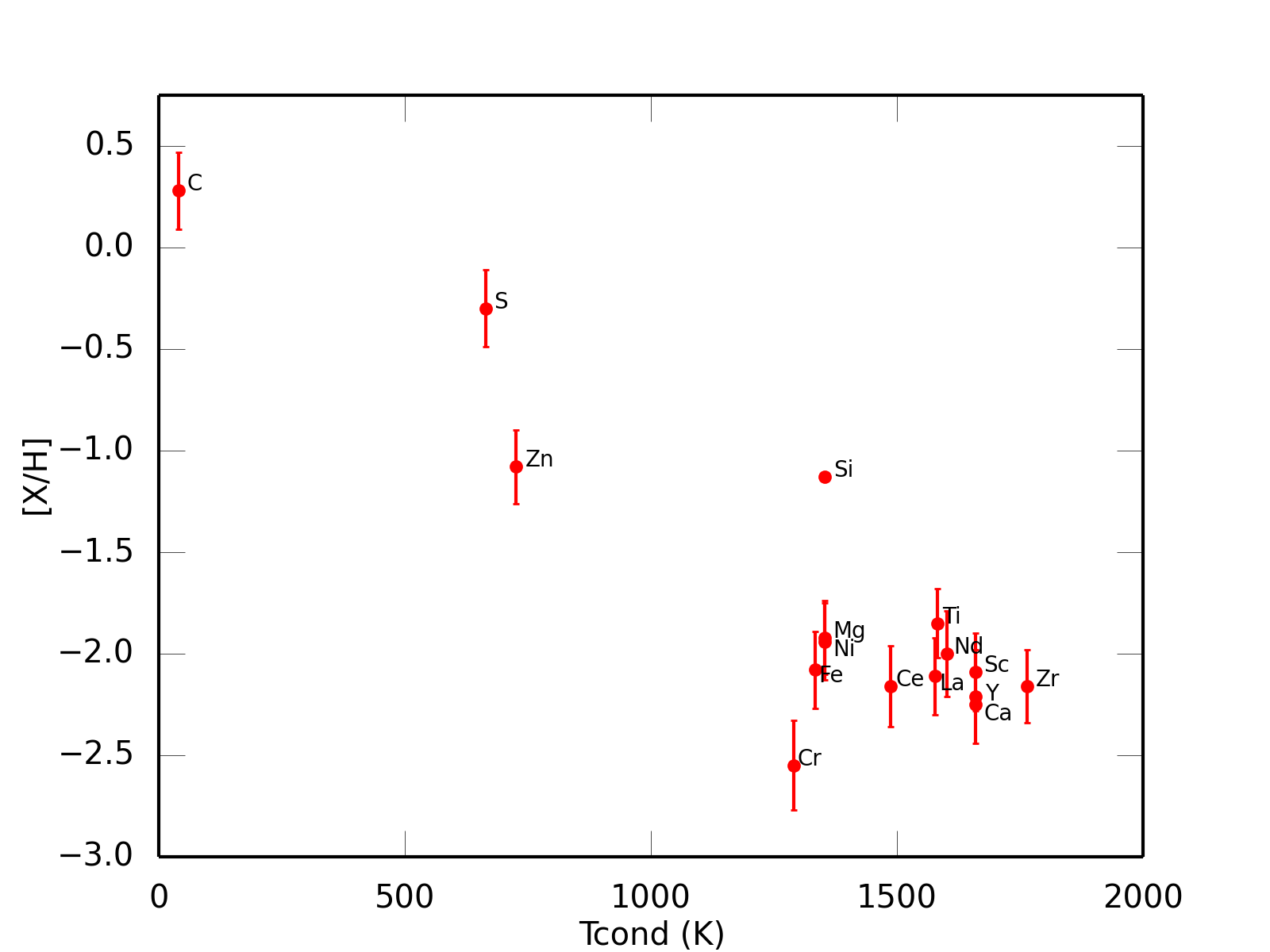}
\caption{The elements abundances in the atmosphere of GZ Nor versus the dust condensation temperature of elements.}\label{fig_7}  
\end{center}
\end{figure}

GZ Nor is a colder star than GK Car. Thus, to find isolated lines with a good line profile has been more difficult. Nevertheless, a sufficient number of good isolated lines (96 in total) have measured 
for abundance analysis. The abundance results are listed in Table \ref {table_7}. The abundance of the elements is plotted versus their dust condensation temperature in Fig. \ref{fig_7}.

GZ Nor is a metal-poor star with [FeI/H] = $-$2.0. We obtained high [S/Fe], [S/Ti], [Zn/Fe] and [Zn/Ti] ([SI/FeI] = +1.7, [IS/TiII] = +1.6, [ZnI/FeI] = +0.7 and [ZnI/TiII] = +0.8) ratios. 
This clearly points to an atmosphere affected by depletion. The amount of \textit {s}$-$process lines are low in GZ Nor's spectra. Nevertheless, we could define abundances 
for Y, Zr, La, Ce, and Nd. The obtained abundances for those \textit {s}$-$process elements are very low, which confirms a depleted photosphere.

\begin{table}
 \centering
  \caption{Abudance analysis of GZ Nor. The solar abundances to compute the [X/H] ratios are taken from
  \protect\cite{asplund09}. The dust condensation temperatures are taken from \protect\cite{lodders09}}.
  \begin{tabular}{lcccc}
  \hline
  & & \multicolumn{1}{c}{GZ Nor}& & \\  
  & & \multicolumn{1}{c}{T$_{eff}$ $=$ 4875 K}& & \\
  & & \multicolumn{1}{c}{log$g$ $=$ 0.5} & & \\
  & & \multicolumn{1}{c}{$\xi_{t}$ $=$ 4.0 km/s}& & \\
  & & \multicolumn{1}{c}{[Fe/H] $=$ $-$2.08}& & \\
 \hline
   ion      & N & [X/H]    & $\sigma$ & T$_{cond}$    \\
  \hline

CI  & 3  &     0.28  & 0.14 & 40  \\
MgI & 3  &    $-$1.92  & 0.16 &1354 \\
SiI & 1  &    $-$1.13  & 0.2  &1354 \\
SI  & 4  &    $-$0.3   & 0.03 &664 \\
CaI & 2  &    $-$2.2   & 0.2  &1659\\
ScII& 4  &    $-$2.09  & 0.18 &1659\\
TiII& 14 &    $-$1.85  & 0.16 &1582\\
CrI & 3  &    $-$2.55  & 0.1  &1291\\
FeI & 27 &    $-$2.08  & 0.11 &1334 \\
FeII& 8  &    $-$1.80  & 0.14 &   \\
NiI & 3  &    $-$1.94  & 0.17 &1353\\
ZnI & 3  &    $-$1.08  & 0.19 &726 \\
YII & 2  &    $-$2.21  & 0.18 &1659\\
ZrII& 4  &    $-$2.16  & 0.12 &1764\\
LaII& 4  &    $-$2.11  & 0.11 &1578\\
CeII& 4  &    $-$2.16  & 0.08 &1487\\
NdII& 5  &    $-$2.0   & 0.14 &1602\\
\hline					        						   	  	  
\end{tabular}
\label{table_7}
\end{table}

\section{Discussion and Conclusion}\label{discussion}

GK Car is a known RV Tauri star and classified as a disc source \citep{evans85, deruyter06, gezer15} in the literature. In this study, thanks to recent ASAS photometry, we show that GZ Nor 
is also an RV Tauri star rather than a long-period variable as was postulated in the literature.

The SED of GZ Nor is classified as an \textquotedblleft uncertain\textquotedblright{} as it is located outside of the disc box as defined in \citep{gezer15}. In this study, we selected
GZ Nor to compare it with a known disc source such as GK Car, which is located in the disc box. Disc sources are generally associated with binarity \citep[see,][]{winckel18}. In our previous study, we 
showed that there is a strong correlation between disc type SED, binarity and the presence of a depleted photosphere \citep{gezer15}. Our abundance analysis shows that both GK Car and GZ Nor display 
depletion of refractory elements. Since the depletion process probably only occurs in a stable disc \citep{waters92} and the disc is associated with binarity \citep{winckel18}, the observed depletion 
pattern in GZ Nor would be an indication of a binary central star surrounded by a circumbinary disc. The dust excess of GZ Nor starts at redder wavelengths than the bulk of the other objects which occupy the 
disc box. The small [3.4]$-$[4.6] colour and the lack of a clear near-IR excess may indicate that the near-IR excess has already disappeared and it's disc is an evolved state. The evolution of post-AGB 
circumbinary discs is not completely understood yet, despite the fact that it has been investigated in many studies \citep[e.g.][]{gielen11, hillen17, bujarrabal13, bujarrabal18}. It is very likely that 
the near-IR luminosity of the disc decreases with time and evolves into a gas-poor debris disc. Hence, the location of these objects in the WISE colour-colour diagram changes according to the evolutionary 
stage of their IR luminosity.

As we discussed in the paper I, the presence of a disc is not a sufficient condition but it is an essential prerequisite for the depletion process to occur. The depletion pattern may still be 
observed while the IR-excess is not detectable anymore \citep{gezer15}. A similar lack of a near-IR excess was observed in AC Her, which is a depleted binary \cite{vanwinckel98}. Its disc was resolved 
using interferometric techniques and it was shown that the circumbinary disc is an evolved state with the large inner radius and low gas/dust ratio \citep{hillen15}. Another example is BD+39$^{\circ}$4926. 
It has long been known that BD+39$^{\circ}$4926 is a strongly depleted object \citep{kodaira70}, but only recently a small IR-excess at 22 $\mu$m detected in W4 band thanks to the sensitivity of the WISE. 
It's binary nature has also been confirmed by \cite{kodaira70} and \cite{gezer15}. Another interesting example is BD+33$^{\circ}$2642, the central star of the PN G052.7+50.7. It has been shown that it 
is a depleted object \citep{napiwotzki94} and a spectroscopic binary \citep{winckel14}, as well. It seems that the observed depletion pattern remains longer than the detectable IR-excess.

\begin{figure}
\begin{center}
%\begin{tabular}{c}
\includegraphics[scale=0.23]{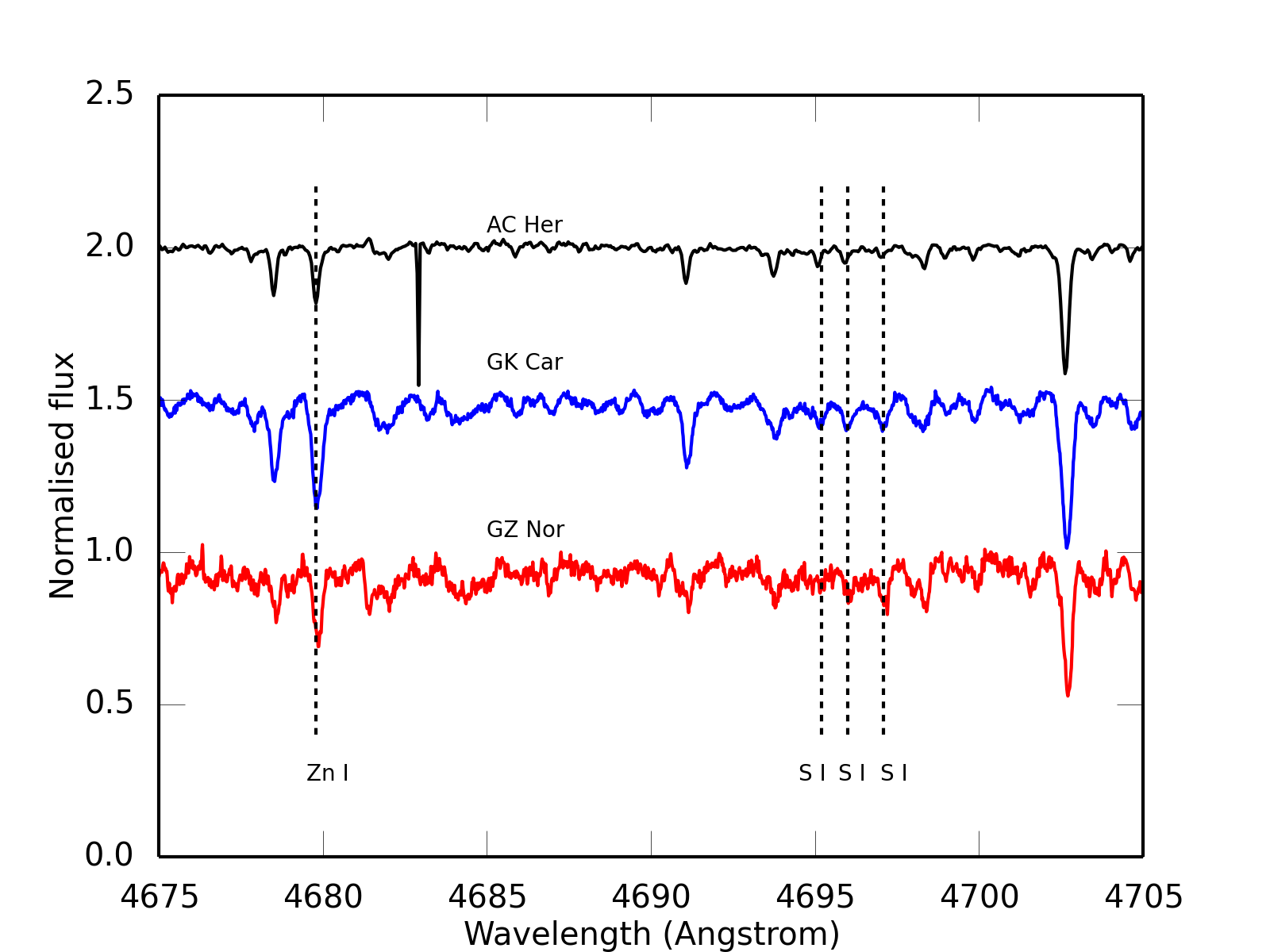}\\
\includegraphics[scale=0.23]{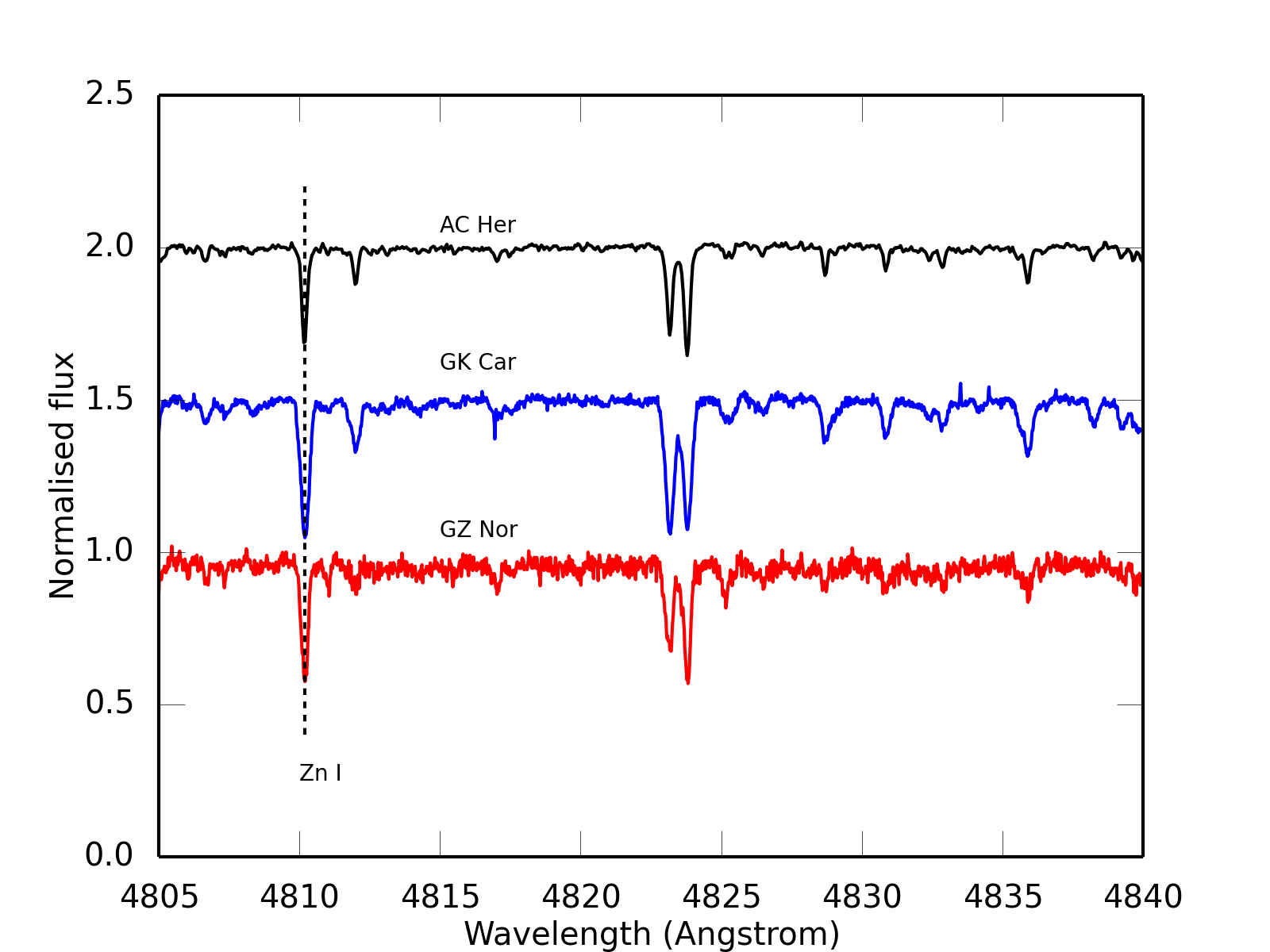}
%\end{tabular}{c}
\caption{S and Zn lines of GK Car and GZ Nor are compared to well-know depleted star AC Her.}\label{fig_8}
\end{center}
\end{figure}

We compare the spectra of GK Car and GZ Nor with the spectra of the well-known binary RV Tauri star AC Her. The orbital period of AC Her is 1194 $\pm{6}$ days \citep{vanwinckel98}. 
AC Her is a depleted RV Tauri star with the disc. Its atmospheric parameters are similar to our stars. In figure \ref{fig_8}, Zn lines around 4680 \AA{} and 4810 \AA{} and S triplet around 4695 \AA{} are 
compared. All three of these stars show depletion in their atmospheres.

The depletion is generally expected to be seen in stars hotter than 5000 K as cooler stars have deep convective envelopes which dilute the accreted gas \citep{giridhar05, venn14, gezer15, oomen18}. 
In this sense, GZ Nor is an interesting example as a cool and depleted object. RV Tauri stars are high-amplitude pulsators and during the pulsation cycle, the temperature could change. 
Their atmosphere states might change dramatically during the pulsation cycle \citep{pollard96, pollard97}. The spectra of GZ Nor was taken at phase $\sim$0.95 during the deep minima. 
This may imply that we took the spectra when the star in the cooler phase.

Finally, since we could constrain the luminosities of our objects, we discuss their evolutionary stages. We compared the observational findings with the recent evolutionary tracks. The derived 
metalicities do not represent the initial metallicities as the atmospheres of the stars are affected by the depletion process. Thus, we have considered evolutionary tracks for a range of metallicities. 
We follow the evolutionary tracks of \cite{bertelli09} for initial masses of 1,2 and 3 M$_{\odot}$ and metallicities of Z=0.002, 0.004 and 0.017. The latter is a solar value.
The luminosities of the tip of the RGB (RGB-tip) for these masses and metallicities are given in Table \ref{table_8}. If we assume that the initial masses of our stars is close to 1 M$_{\odot}$, 
in that case, their luminosities obtained from different methods are lower than the expected RGB-tip luminosities for all three different metallicities. However, if we assume that the initial masses of 
our stars is 2 or 3 M$_{\odot}$, then their luminosities are higher than the RGB-tip luminosities of 2, 3 M$_{\odot}$ stars, for all three different metallicities. Therefore, our stars could be
post-RGB objects, which are progenitors of 1 M$_{\odot}$ star or post-AGB objects, which are progenitors of 2 or 3 M$_{\odot}$ stars. One should note that uncertainties are large in the derived 
luminosities as it is seen in tables \ref{table_4} and \ref{table_5}. Considering their disc-type SEDs and the depletion characteristics, we are assuming that they are indeed binary objects. 
Thus, they might have evolved off the RGB or AGB due to a binary interaction process.

\begin{table}
%  \setlength{\tabcolsep}{1pt}
%    \scriptsize
    \caption{Luminosities at the tip of the RGB from evolutionary tracks of \protect\cite{bertelli09} for a range of initial masses and metallicities.}
      \begin{tabular}{cccc}

       \hline  
         mass     &Z = 0.017 & Z = 0.004 & Z = 0.002 \\
        \hline
              
  1M$_{\odot}$&3160L$_{\odot}$& 2880L$_{\odot}$ &2640L$_{\odot}$ \\
  2M$_{\odot}$&855L$_{\odot}$ & 335L$_{\odot}$  &310L$_{\odot}$  \\
  3M$_{\odot}$&540L$_{\odot}$ & 665L$_{\odot}$  &690L$_{\odot}$  \\
  \hline                                                                  
\end{tabular}
\label{table_8}
\end{table}

In this study, we presented a photometric and spectroscopic analysis of two RV Tauri stars GK Car and GZ Nor. Thanks to recent ASAS photometry, we showed that GZ Nor is an RV Tauri star. Using 
the high-resolution UVES spectra, we performed a detailed abundance analysis and showed that our two stars display depletion of refractory elements and this again confirm the idea that depletion is 
common in the disc sources and the RV Tauri stars with the disc-type SED are likely all binaries \citep{manick17}.  We defined the luminosities and distances, which is important to understand their 
evolutionary status. They are both dusty objects with relatively low luminosity, they could be either post-AGB or post-RGB objects \citep{kamath15}.

\section*{Acknowledgements}

This work has been performed thanks to 2214-A International Research Fellowship Programme of the Scientific and Technological Research Council of Turkey (TUBITAK). 

HVW acknowledged the support of the KU Leuven research council (contract C14/17/082). IG would like to thank KU Leuven Astronomy Institute for their kind hospitality.

IG acknowledges her thanks to Prof. Ryszard Szczerba for his comments and suggestions.

Based on observations obtained with the UVES spectrograph of ESO on proposal 074.D-0619 (PI Van Winckel Hans).

The following Internet-based resources were used in research for this paper: the NASA Astrophysics Data System; the SIMBAD data
base and the VizieR service operated by CDS, Strasbourg, France.

%%%%%%%%%%%%%%%%%%%%%%%%%%%%%%%%%%%%%%%%%%%%%%%%%%



% \bibliographystyle{mnras}
\bibliography{mnemonic,gezer}

% % Alternatively you could enter them by hand, like this:
% % This method is tedious and prone to error if you have lots of references
% \begin{thebibliography}{99}
% \bibitem[\protect\citeauthoryear{Author}{2012}]{Author2012}
% Author A.~N., 2013, Journal of Improbable Astronomy, 1, 1
% \bibitem[\protect\citeauthoryear{Others}{2013}]{Others2013}
% Others S., 2012, Journal of Interesting Stuff, 17, 198
% \end{thebibliography}

%%%%%%%%%%%%%%%%%%%%%%%%%%%%%%%%%%%%%%%%%%%%%%%%%%

%%%%%%%%%%%%%%%%% APPENDICES %%%%%%%%%%%%%%%%%%%%%

\appendix

% \section{Some extra material}
% 
% If you want to present additional material which would interrupt the flow of the main paper,
% it can be placed in an Appendix which appears after the list of references.

%%%%%%%%%%%%%%%%%%%%%%%%%%%%%%%%%%%%%%%%%%%%%%%%%%

% Don't change these lines
\bsp	% typesetting comment
\label{lastpage}
\end{document}